\documentclass[prd,showpacs,showkeys,floatfix,amsmath,amssymb,floatfix]{revtex4}
\usepackage{graphicx,color,dcolumn,booktabs,bm}
\usepackage{longtable,lscape}
\usepackage{amssymb}
\usepackage{indentfirst}
\usepackage{epsfig}

\begin{document}

\title{Can  $X(5568)$ be a tetraquark state?}
\author{Wei Wang}
 \email{wei.wang@sjtu.edu.cn}
\author{Ruilin Zhu\footnote{Corresponding author}}
\email{rlzhu@sjtu.edu.cn}

\affiliation{
INPAC, Shanghai Key Laboratory for Particle Physics and Cosmology, Department of Physics and Astronomy, Shanghai Jiao Tong University, Shanghai 200240,  China and\\
State Key Laboratory of Theoretical Physics, Institute of Theoretical Physics, Chinese Academy of Sciences, Beijing 100190, China
}

\begin{abstract}
Very recently, the D0 collaboration has reported the observation of a narrow structure, $X(5568)$, in the decay process  $X(5568)\to  B^0_s\pi^\pm$ using  the 10.4${\rm fb}^{-1}$ data of $p\bar p$ collision at $\sqrt s= 1.96$ TeV.  This structure is of great interest since it is the first hadronic state with  four different valence quark flavors, $b,s,u,d$.
In this work, we investigate  tetraquarks with four different quark flavors.  Based on the diquark-antidiquark scheme, we study the spectroscopy of the tetraquarks with one heavy bottom/charm quark and three light quarks. We find that the lowest-lying  S-wave   state, a tetraquark with the flavor  $[su][\bar b\bar d ]$ and the spin-parity $J^P=0^+$,  is about 150 MeV higher than the $X(5568)$.   Further detailed experimental and theoretical  studies of the spectrum, production and decays of  tetraquark states with four different flavors  in the future are severely needed  towards  a better understanding  its  nature and the classification  of hadron exotic states.
\end{abstract}


\maketitle

Since the proposal of the concept of quarks by Gell-Mann~\cite{GellMann:1964nj}, there have been great endeavors to  test the quark model and  search for exotic structures beyond  this scheme.  To date hundreds of   hadrons were
discovered, and most  of them can be accommodated  in the naive quark model, in which  mesons and
baryons are composed of a quark--antiquark pair and three quarks, respectively. No  firm evidence for the existence of   exotic states beyond the quark model has been established on experimental side until the discovery of the $X(3872)$ in 2003~\cite{Choi:2003ue,Acosta:2003zx,Aubert:2004ns,Abazov:2004kp}.  The peculiar properties of the $X(3872)$, the vicinity of its mass close to $D\bar D^*$ threshold, the tiny width and the large isospin violation in its production and decay,  has    invoked a renaissance of hadron spectroscopy studies. Since then one key topic in hadron physics is the identification of the exotic hadrons. Many new interesting structures were discovered in the mass region of heavy
quarkonium, named as  $XYZ$ states (for a review  of these particles, see Refs.~\cite{Brambilla:2010cs,Agashe:2014kda,Esposito:2014rxa,Chen:2016qju}). In particular,  the charged structures with a hidden
pair of heavy quark and antiquark such as the
$Z_c^\pm(4430)$~\cite{Choi:2007wga,Aaij:2014jqa},
$Z_b^\pm(10610,10650)$~\cite{Belle:2011aa},
$Z_c^\pm(3900)$~\cite{Ablikim:2013mio,Liu:2013dau}, and
$Z_c^\pm(4020)$~\cite{Ablikim:2013emm} would be undoubtedly  exotic resonances.   Moreover  candidates for
exotic hadrons were  extended to the pentaquark sector by the LHCb
observations of two structures in the $J/\psi\, p$ invariant
mass distribution with masses (widths) $(4380\pm8\pm29)$~MeV
($(205\pm18\pm86)$~MeV) and  $(4449.8\pm1.7\pm2.5)$~MeV ($(39\pm5\pm19)$~MeV),
respectively~\cite{Aaij:2015tga}.  These discoveries have opened up a new era of multi-quark spectroscopy  and strengthened our belief  that the hadron spectrum would  be much richer than the quark model.

Most of the observed exotic $X,Y,Z$ structures  so far share a common feature, i.e.  they consist of a hidden heavy quark-antiquark pair,  $\bar bb$ or $\bar cc$.  Various theoretical models were motivated   to explain these exotics, many of which have made use of the heavy quark symmetry and chiral symmetry. Inspired by these symmetries,  various combinations of heavy and light mesons have been examined and can be  searched for, of great interest is the one that is composed of a heavy meson and a light meson.  Very recently,  the D0 collaboration has reported the first observation of such structure in the final state $B_s^0\pi^\pm$~\cite{D0:2016mwd}. Since the  $B_s^0\pi^\pm$ final state is made of four different flavors, $b,s,u,d$, the new structure is definitely exotic.  Its mass  and  width has been determined as~\cite{D0:2016mwd}
\begin{eqnarray}
 M_X= (5567.8\pm2.9){\rm MeV}, \;\;\; \Gamma_X= (21.9\pm 6.4){\rm MeV}.\label{eq:X5569_D0}
\end{eqnarray}
The above results are obtained through a fit based on  a Breit-Wigner parametrization  of the S-wave decay of $X(5568)\to B_s^0\pi^\pm$.
The statistical significance including the look-elsewhere effect and systematic errors is about $5.1\sigma$~\cite{D0:2016mwd}.

After the first discovery,  more experimental efforts to determine the properties of the $X(5568)$ are needed . Meanwhile,  this also requests theoretical interpretations of its nature.
The $X(5568)$ is too far from the the $B_d^0K^\pm$ threshold (5774 MeV) to be interpreted as a hadronic molecule of $B_d^0 K^\pm$. In addition,  the interaction of $B_s^0\pi^\pm$ is very weak and unable to form a bounded structure~\cite{Guo:2009ct}. In this work,  we will study tetraquark states using  colored components, diquark and antidiquark,  and bound by the long-range color forces.  Tetraquark states in the large $N_c$ limit of QCD has been explored  in Refs.~\cite{Weinberg:2013cfa,Knecht:2013yqa,Lebed:2013aka}, which indicates that a compact tetraquark meson may have narrow decay widths scale as $1/N_c$. Thereby there are reasonable candidates for the additional spectroscopic hadron series apart from the quark model. We want to see whether the $X(5568)$ is a tetraquark state.  In the past decades,  tetraquark states in particular with hidden bottom and charm have been explored in Refs.~\cite{Chen:2004dy,Cheng:2003kg,Maiani:2004vq,Ali:2009pi,Ali:2009es,Ali:2010pq,Ali:2011ug,Ali:2014dva,Liu:2014dla,
Maiani:2014aja,Ma:2014zva,Wang:2016mmg,Lebed16} and many references therein.

The  QCD confining potential for the multiquark system can be generally written as~\cite{DeRujula:1975qlm}
\begin{eqnarray}
V(\vec{r}_i) &=& L(\vec{r}_1,\vec{r}_2,\ldots)+\sum_{i>j}I\, \alpha_s S_{ij}\ ,
\end{eqnarray}
where the $L(\vec{r}_1,\vec{r}_2,\ldots)$ stands for  the universal binding interaction of quarks. The $S_{ij}$ is two-body Coulomb and chromomagnetic interactions, with  the $I=-4/{3}$ and $-2/3$ as  the single gluon interaction strength in quark-antiquark and quark-quark cases, respectively.

In the following, we will consider the tetraquark states  with quark content $[qq^\prime][\bar{q}\bar{Q}]$, where $q$ and $q^\prime$ denote the light quarks, and $Q$ denotes a  heavy quark,   bottom or charm. The effective Hamiltonian is composed of three kinds of interactions: spin-spin interactions of   quarks in  the diquark and antidiquark, and  between them; the spin-orbital interaction; purely orbital interactions. An explicit model that incorporates these interactions has been established  in Ref.~\cite{Maiani:2004vq}:
\begin{eqnarray}
 H&=&m_{\delta}+m_{\delta^\prime}+H^{\delta}_{SS} + H^{\bar{\delta^\prime}}_{SS}+H^{\delta\bar{\delta^\prime}}_{SS} + H_{SL}+H_{LL},\nonumber\\
 \label{eq:definition-hamiltonian}
\end{eqnarray}
with the functions
\begin{eqnarray}
 H^\delta_{SS}&=&2(\kappa_{q q^\prime})_{\bar{3}}(\mathbf{S}_q\cdot \mathbf{S}_{q^\prime}),\nonumber\\
 H^{\bar{\delta^\prime}}_{SS}&=&2(\kappa_{ Qq})_{\bar{3}}(\mathbf{S}_{\bar{Q}}\cdot \mathbf{S}_{\bar{q}}), \nonumber\\
 H^{\delta\bar{\delta^\prime}}_{SS} &=&2\kappa_{q\bar{q}}(\mathbf{S}_q\cdot \mathbf{S}_{\bar{q}}) +2\kappa_{q^{\prime} \bar{q}}(\mathbf{S}_{q^{\prime}}\cdot \mathbf{S}_{\bar{q}})\nonumber\\&&+2\kappa_{q\bar{Q}} (\mathbf{S}_q\cdot \mathbf{S}_{\bar{Q}})+ 2\kappa_{q^\prime\bar{Q}}(\mathbf{S}_{q^\prime}\cdot \mathbf{S}_{\bar{Q}}),
\nonumber\\
 H_{SL}&=&2 A_\delta (\mathbf{S}_\delta \cdot \mathbf{L}) +2 A_{\bar{\delta^\prime}}(\mathbf{S}_{\bar{\delta^\prime}}\cdot \mathbf{L}),\nonumber\\
 H_{LL}&=&B_{\delta\bar{\delta^\prime}} \frac{L(L+1)}{2}\ .
\label{eq:definition-hamiltonian2}
\end{eqnarray}
In the above, the $m_\delta$ and $m_{\delta^\prime}$ is the constituent mass of the diquark $[qq^\prime]$ and the antidiquark $[\bar{q}\bar{Q}]$, respectively.  The spin-spin interaction inside the diquark and antidiquark is denoted as   $ H^\delta_{SS}$ and $H^{\bar{\delta^\prime}}_{SS}$, respectively. The $H^{\delta\bar{\delta^\prime}}_{SS}$ reflects  the spin-spin interaction of quarks between diquark and antidiquark. The  $H_{SL}$ and $H_{LL}$ is the spin-orbital and purely orbital terms.  The $\mathbf{S}_{\delta}$ and $\mathbf{S}_{\bar{\delta^\prime}}$ corresponds to the spin operator  of diquark and antidiquark, respectively. The spin operator  of light quarks and  heavy antiquark  is given by  $\mathbf{S}_{q^{(\prime)}}$ and $\mathbf{S}_{\bar{Q}}$, respectively.  The symbol  $\mathbf{L}$ denotes  the orbital angular momentum operator.  The coefficients $\kappa_{q_1\bar{q}_2}$ and $(\kappa_{q_1 q^\prime_2})_{\bar{3}}$ are the spin-spin couplings for a quark-antiquark pair and diquark in color antitriplet, respectively; $A_{\delta(\bar{\delta^\prime})}$ and $B_{\delta\bar{\delta^\prime}}$ denotes respectively spin-orbit and orbit-orbit couplings.

For the lowest-lying  tetraquark states with  the quark content $[qq^\prime][\bar{q}\bar{Q}]$, their orbital angular momenta are vanishing, i.e. $L=0$.  Among them,  there are two possible tetraquark configurations  with the spin-parity $J^P=0^+$, i.e.,
\begin{eqnarray}
|0_J\rangle_1&=&\frac{1}{2}
\big[(\uparrow)_q(\downarrow)_{q^\prime}-(\downarrow)_q(\uparrow)_{q^\prime} \big]\big[(\uparrow)_{\bar{q}}(\downarrow)_{\bar{Q}}
-(\downarrow)_{\bar{q}}(\uparrow)_{\bar{Q}}\big],\nonumber\\
|0_J\rangle_2&=&\frac{1}{\sqrt{3}}
\big\{(\uparrow)_q(\uparrow)_{q^\prime}(\downarrow)_{\bar{q}}(\downarrow)_{\bar{Q}}
+(\downarrow)_q(\downarrow)_{q^\prime}(\uparrow)_{\bar{q}}(\uparrow)_{\bar{Q}} \nonumber\\&& -\frac{1}{2}
\big[(\uparrow)_q(\downarrow)_{q^\prime}+(\downarrow)_q(\uparrow)_{q^\prime} \big](\uparrow)_{\bar{q}}(\downarrow)_{\bar{Q}}
\nonumber\\&& -\frac{1}{2}
\big[(\uparrow)_q(\downarrow)_{q^\prime}+(\downarrow)_q(\uparrow)_{q^\prime} \big](\downarrow)_{\bar{q}}(\uparrow)_{\bar{Q}}\big\}.
 \label{eq:definition-states0+}
\end{eqnarray}
In the above, $|0_J\rangle_1=|0_\delta,0_{\bar{\delta^\prime}},0_J\rangle$, $|0_J\rangle_2=|1_\delta,1_{\bar{\delta^\prime}},0_J\rangle$, and $|S_\delta,S_{\bar{\delta^\prime}},S_J\rangle $ stands for the tetraquark; the $S_\delta$ and $S_{\bar{\delta^\prime}}$ stand for  the spin  of diquark $[qq^\prime]$ and antidiquark $[\bar{q}\bar{Q}]$, respectively, while the $S_J$ denotes the total angular momentum of the tetraquark.
In this paper,  we only  focus on the scalar and vector diquarks, i.e. $S_{\delta^{(\prime)}}=0,1$.

Using the basis   defined in Eq. (\ref{eq:definition-states0+}), one can derive  the mass matrix for the $J^P=0^+$ tetraquarks

\begin{widetext}
\begin{eqnarray}
M= m_{\delta}+m_{\delta'}+ \left(
\begin{array}{cc}
 -\frac{3}{2} ((\kappa_{q{q^\prime}})_{\bar{3}}+(\kappa_{Q q})_{\bar{3}}) & \frac{\sqrt{3}}{2}  (\kappa_{q^\prime \bar{Q}}+\kappa_{q\bar{q}}-\kappa_{q^\prime \bar{q}}-\kappa_{q\bar{Q}}) \\
 \frac{\sqrt{3}}{2}  (\kappa_{q^\prime \bar{Q}}+\kappa_{q\bar{q}}-\kappa_{q^\prime \bar{q}}-\kappa_{q\bar{Q}}) & \frac{1}{2} ((\kappa_{q{q^\prime}})_{\bar{3}}+(\kappa_{Q q})_{\bar{3}}-2 \kappa_{q^\prime \bar{q}}-2 \kappa_{q^\prime \bar{Q}}-2 \kappa_{q\bar{q}}-2 \kappa_{q\bar{Q}})
\end{array} \right).
\end{eqnarray}
\end{widetext}

In the case  $J^P=1^+$, there are three possible tetraquark states, i.e.,
\begin{eqnarray}
|0_\delta,1_{\bar{\delta^\prime}},1_J\rangle&=&\frac{1}{\sqrt{2}}
\big[(\uparrow)_q(\downarrow)_{q^\prime}-(\downarrow)_q(\uparrow)_{q^\prime} \big](\uparrow)_{\bar{q}}(\uparrow)_{\bar{Q}}
 ,\nonumber\\
 |1_\delta,0_{\bar{\delta^\prime}},1_J\rangle&=&\frac{1}{\sqrt{2}}
(\uparrow)_q(\uparrow)_{q^\prime}\big[(\uparrow)_{\bar{q}}(\downarrow)_{\bar{Q}}-(\downarrow)_{\bar{q}}
(\uparrow)_{\bar{Q}}\big]
 ,\nonumber\\
|1_\delta,1_{\bar{\delta^\prime}},1_J\rangle&=&\frac{1}{2}
\big\{(\uparrow)_q(\uparrow)_{q^\prime}\big[(\uparrow)_{\bar{q}}(\downarrow)_{\bar{Q}}+(\downarrow)_{\bar{q}}
(\uparrow)_{\bar{Q}}\big]\nonumber\\&&-\big[(\uparrow)_q(\downarrow)_{q^\prime}+(\downarrow)_q(\uparrow)_{q^\prime} \big](\uparrow)_{\bar{q}}(\uparrow)_{\bar{Q}}\big\}.
 \label{eq:definition-states1+}
\end{eqnarray}
Notice that unlike the heavy quarkonium-like states, the tetraquarks with the quark content $[qq^\prime][\bar{q}\bar{Q}]$ do not have any definite charge parity and thus the above three $1^+$ states can mix with each other.

Using the basis   defined in Eq. (\ref{eq:definition-states1+}), one can obtain  the mass splitting matrix $\Delta M$ for $J^P=1^+$
\begin{widetext}
\begin{eqnarray}
\Delta M=\left(
\begin{array}{ccc}
 \frac{1}{2} ((\kappa_{Q q})_{\bar{3}}-3 (\kappa_{q{q^\prime}})_{\bar{3}}) & \frac{1}{2} (\kappa_{q^\prime \bar{Q}}-\kappa_{q^\prime \bar{q}}-\kappa_{q\bar{Q}}+\kappa_{q\bar{q}}) & \frac{\sqrt{2}}{2}(\kappa_{q\bar{Q}}-\kappa_{q^\prime \bar{Q}}-\kappa_{q^\prime \bar{q}}+\kappa_{q\bar{q}}) \\
 \frac{1}{2} (\kappa_{q^\prime \bar{Q}}-\kappa_{q^\prime \bar{q}}-\kappa_{q\bar{Q}}+\kappa_{q\bar{q}}) & \frac{1}{2} ((\kappa_{q{q^\prime}})_{\bar{3}}-3 (\kappa_{Q q})_{\bar{3}}) & \frac{\sqrt{2}}{2}(\kappa_{q^\prime \bar{Q}}-\kappa_{q^\prime \bar{q}}+\kappa_{q\bar{Q}}-\kappa_{q\bar{q}}) \\
 \frac{\sqrt{2}}{2}(\kappa_{q\bar{Q}}-\kappa_{q^\prime \bar{Q}}-\kappa_{q^\prime \bar{q}}+\kappa_{q\bar{q}}) & \frac{\sqrt{2}}{2}(\kappa_{q^\prime \bar{Q}}-\kappa_{q^\prime \bar{q}}+\kappa_{q\bar{Q}}-\kappa_{q\bar{q}}) & \frac{1}{2}
   ((\kappa_{q{q^\prime}})_{\bar{3}}+(\kappa_{Q q})_{\bar{3}}-\kappa_{q^\prime \bar{Q}}-\kappa_{q^\prime \bar{q}}-\kappa_{q\bar{Q}}-\kappa_{q\bar{q}})
\end{array}
\right),\nonumber
\end{eqnarray}
\end{widetext}
and the mass matrix is given as
\begin{eqnarray}
 M= m_{\delta}+m_{\delta'}+\Delta M.
\end{eqnarray}

For the $J^P=2^+$, there exits only one tetraquark configuration:
\begin{eqnarray}
|1_\delta,1_{\bar{\delta^\prime}},2_J\rangle&=&
(\uparrow)_q(\uparrow)_{q^\prime}(\uparrow)_{\bar{q}}(\uparrow)_{\bar{Q}},
\label{eq:definition-states2+}
\end{eqnarray}
with the mass
\begin{eqnarray}
 M(2^+)&=&m_\delta+m_{\delta^\prime}+\frac{1}{2}\left( (\kappa_{q{q^\prime}})_{\bar{3}}+(\kappa_{Q q})_{\bar{3}}\right)+\frac{1}{2}\left(\kappa_{q\bar{q}}+\kappa_{q\bar{Q}}+\kappa_{q^\prime \bar{q}}+\kappa_{q^\prime \bar{Q}}\right).
\end{eqnarray}
The spin-spin couplings have been extensively  explored  in the  previous analyses of  mesons, baryons and the $XYZ$ spectra in quark model and diquark model. We quote the results from Refs.~\cite{Maiani:2004vq,Ali:2009es,Ali:2011ug,Ali:2014dva} and summarize  them in Table~\ref{tab:spin-spin coupling}.
The masses of diquarks $[cq]$ and $[bq]$ are determined through the analysis of the $X(3872)$ with $J^{PC}=1^{++}$ and $Y_b(10890)$ with $J^{PC}=1^{--}$
in the diquark model, respectively. We quote $m_{[cq]}=1.932${GeV} and $m_{[bq]}=5.249${GeV} \cite{Ali:2009es,Ali:2011ug,Zhu:2015bba}.
 Using these results for  the spin-spin, spin-orbit and orbit-orbit couplings and diquark masses, we can  obtain the tetraquark spectrum.  The tetraquark spectra with quantum number $J^P=
0^+$, $1^+$, and $2^+$ are depicted in Fig.~\ref{Fig-spectrum1}, in which the left and right panel corresponds to the tetraquark with a bottom and a charm quark respectively.  The masses given in the figure are in units of GeV.  The thresholds of the $B_s\pi, B_s^*\pi, B_{s2}\pi$ and their charm analogues are shown in dashed lines.  The masses of the $X(5568)$ and the $D_{s0}^*(2317)$ are also given in the figure.

\begin{widetext}
\begin{table}[thb]
\caption{\label{tab:spin-spin coupling} The spin-spin couplings (in units of MeV) for color-singlet quark-antiquark and color-antitriplet quark-quark pairs. The relation  $\kappa _{ij}=(\kappa _{ij})_{0}/4$ for quark-antiquark coupling is obtained in one gluon exchange model. Here  the $q$ denotes the light  $u$ and $d$ quarks, where we have adopted the isospin symmetry.}
\begin{center}
\begin{tabular}{c|cccccccc||c|cccccc}
\hline\hline
 Quark-antiquark &$q\bar{q}$&$s\bar{s}$&$s\bar{q}$&$c\bar{q}$&$c\bar{s}$&$c \bar{c}$&$b\bar{q}$&$b\bar{s}$&Diquark &qq&ss&sq&cq&cs&bq
 \\
 Couplings $(\kappa _{ij})_{0}$&315&121&195&70&72&59&23&23&
 Couplings $(\kappa _{ij})_{\bar{3}}$&103&72&64&22&25&6.6\\
\hline\hline
\end{tabular}
\end{center}
\end{table}
\begin{figure}[th]
\begin{center}
\includegraphics[width=0.7\textwidth]{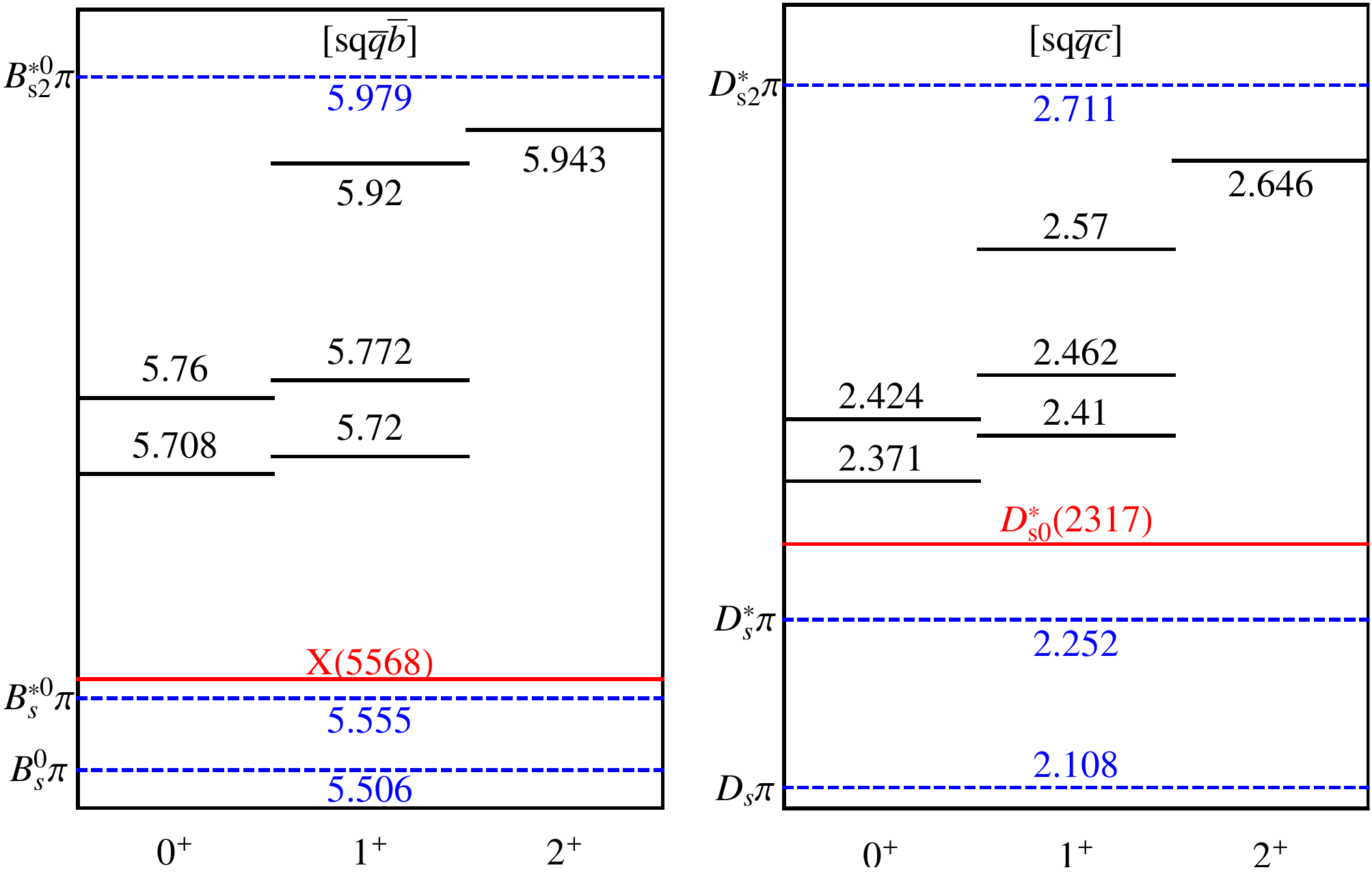}
\end{center}
\vskip -0cm
\caption{The open bottom (left panel) and charm (right panel) tetraquark spectra in the diquark-antidiquark scheme.  The masses given in the panels are in units of GeV.  The thresholds of the $B_s\pi, B_s^*\pi, B_{s2}\pi$ and their charm analogues are shown in dashed lines.  The masses of the $X(5568)$ and the $D_{s0}^*(2317)$ are also given in the figure.  }
\label{Fig-spectrum1}
\end{figure}
\end{widetext}

Our results for tetraquarks with a charm quark in Fig.~\ref{Fig-spectrum1} are consistent with Ref.~\cite{Maiani:2004vq}, in which we find that the lowest tetraquark state is about $60$ MeV higher than the discovered  $D_{s0}^*(2317)$. Switching to the bottom sector, we find this difference gets bigger: our prediction for the mass of the lower S-wave $0^+$ tetraquark  is about 150 MeV larger than the experimental result by D0 for the mass of the $X(5568)$ in Eq.~\eqref{eq:X5569_D0}. The deviation on mass of the $X(5568)$ still exists when reducing the diquark masses in a reasonable region.     In the charm sector, the $D_{s0}^*(2317)$ is about 70 MeV higher than the $D_s^*\pi$ threshold, while in the bottom sector, the observed $X(5568)$ by D0 is only about 10 MeV higher than the $B_s^*\pi$ threshold. In the heavy quark limit, the mass difference between the lowest tetraquark state  and the  $D_s^*/B_s^*\pi$  presumably arises from   the excitation of the light system, and  might be at the same magnitude  in the bottom sector and charm sector. Thus data  may indicate that the $X(5568)$ is too light to be the partner of the $D_{s0}^*(2317)$.  In order to simultaneously describe the tetraquarks with four different flavors, a more comprehensive analysis is called for  in future. In addition, the open bottom (charm) tetraquark states with strangeness number $S=-1$ constitute an isospin triplet and a  singlet. Therein the tetraquark states with the quark content $[su][\bar b\bar d ]$, $[sd][\bar b\bar u ]$ and $1/{\sqrt{2}}([su][\bar b\bar u ]-[sd][\bar b\bar d ])$ constitute an isospin triplet, while the tetraquark with $1/{\sqrt{2}}([su][\bar b\bar u ]+[sd][\bar b\bar d ])$ is an isospin singlet. Due to the isospin symmetry, the masses of these tetraquark partners  are identical to  the values given in Fig.~\ref{Fig-spectrum1}.

In the past decades,  the spectroscopy  study of hadron exotics has played an important role in uncovering the hadron inner structure, and  examining various models for hadrons with fundamental freedom. Many of the recently observed structures defy an ordinary interpretation as a $\bar qq$ meson or a $qqq$ baryon.   In this work, we have explored   the tetraquarks with four different quark flavors.  Based on the diquark-antidiquark scheme, we have calculated  the spectroscopy of the tetraquarks with one heavy bottom/charm quark and three light quarks. We find that the lowest-lying  S-wave   state, a  $J^P=0^+$ tetraquark with the flavor   $[su][\bar b\bar d ]$, lies at around $5.7$GeV and   is about 150 MeV higher than the $X(5568)$. The identification of the $X(5568)$ as a tetraquark with spin-parity $J^P=0^+$ is a challenge to the tetraquark model.  Actually,
the LHCb Collaboration did not see  a signal for the $X(5568)$, where the invariant mass of $B_s^0\pi^\pm$ is scanned between $5.5$GeV and $5.7$GeV~\cite{LHCb:2016ppf}. So it is worth to search tetraquark states with four different flavors in the invariant mass of $B_s^0\pi^\pm$  beyond $5.7$GeV.
Further detailed experimental and theoretical  studies of the spectrum, production and decays of  tetraquark states with four different flavors in the future are severely called for  towards  a better understanding  its  nature and the classification  of hadron exotics.
\vspace{1mm}

\emph{Note added:}After this work was finished,   a series of papers also investigated the structure of the $X(5568)$,
where a diquark-antidiquark interpretation is employed in Refs.~\cite{Agaev:2016mjb,Chen:2016mqt,Wang:2016mee,Zanetti:2016wjn,Liu:2016ogz,Tang:2016pcf}; threshold rescattering effect and is studied in Ref.~\cite{Liu:2016xly}, and Ref.~\cite{Xiao:2016mho}, respectively; decay relations in flavor SU(3) symmetry  is studied in Ref.~\cite{He:2016yhd}.

\vspace{2mm}
\acknowledgments{The authors thank   Feng-Kun Guo,   Xiangdong Ji, Gang Li,  Xiao-Hai Liu, Cong-Feng Qiao and Zhen-Jun Xiao for valuable discussions. W. Wang  has benefited a lot from the collaboration with Ahmed Ali,  Christian Hambrock and Satoshi Mishima  on the tetraquark model. 
This work is supported by National Natural Science
Foundation of China (No.11575110), by Natural  Science Foundation of Shanghai under Grant  No. 15DZ2272100 and No. 15ZR1423100,  by China Postdoctoral Science Foundation, by the Open Project Program of State Key Laboratory of Theoretical Physics, Institute of Theoretical Physics, Chinese  Academy of Sciences, China (No.Y5KF111CJ1), and  by   Scientific Research Foundation for   Returned Overseas Chinese Scholars, State Education Ministry.}

\end{document}